\documentstyle[aps,twocolumn,prl,epsfig]{revtex}

\def\be{\begin{equation}}
\def\ee{\end{equation}}
\def\ba{\begin{eqnarray}}
\def\ea{\end{eqnarray}}
\def\bq{\begin{quote}}
\def\eq{\end{quote}}

\newcommand{\gsim}{\lower.7ex\hbox{$\;\stackrel{\textstyle>}{\sim}\;$}}
\newcommand{\lsim}{\lower.7ex\hbox{$\;\stackrel{\textstyle<}{\sim}\;$}}
\begin{document}
\twocolumn[\hsize\textwidth\columnwidth\hsize\csname @twocolumnfalse\endcsname

\title{ Unifying the Strengths of Forces in Higher Dimensions}

\author{A. Dedes$^1$ and P. Kanti$^2$}

\address{$^1$Rutherford Appleton Laboratory, Chilton, Didcot,
Oxon OX11 OQX, UK\\
$^2$Theoretical Physics Institute, School of Physics
and Astronomy,\\ University of Minnesota, Minneapolis, MN 55455, USA}

\maketitle

\begin{abstract}
We consider the embedding of the Standard Model fields in a
$(4+d)$-dimensional theory while gravitons may propagate in $d'$
extra, compact dimensions. We study the modification of strengths
of the gravitational and gauge interactions and, for various values
of $d$ and $d'$, we determine the energy scale at which these
strengths are unified. Special cases where the unification 
of strengths is characterized by the absence of any hierarchy
problem are also presented.
\end{abstract}
\vspace*{3mm}
]

It is widely believed that any fundamental theory capable of
describing our world at higher energy scales always predicts the
existence of extra, spatial, compact dimensions. Being motivated by
attempts to lower the string scale at the gauge unification
scale~\cite{witten} or the TeV scale~\cite{Lykken},
the concept of large~\cite{ADD} or
small~\cite{RS} extra dimensions, that are being felt only by
gravitons, has been used in order to attack the hierarchy problem.
On the other hand, the existence of extra dimensions,
where gauge bosons can propagate, was
proposed in an attempt to explain the size of the supersymmetry
breaking scale~\cite{antoniadis} or to lower the unification scale
of the gauge interactions~\cite{unification1}. 
In both cases, the Kaluza-Klein excitations of either gravitons
or Standard Model particles appear in the framework of the
4-dimensional, effective theory and, in principle, may modify
the low-energy physics. 
The accuracy with which the electroweak and strong interactions
have been probed at low energies demands that any contributions
to SM processes coming from the KK excitations be extremely
small which, in turn, places a lower bound on the energy scale
associated with the extra dimensions.

In this letter, we assume the existence of two different sets of
extra dimensions: one being felt only by gravitons and one where
gauge bosons can propagate. The presence of the extra
dimensions modifies simultaneously the strength of both the gravitational
and gauge interactions. We search for specific combinations of
the numbers and sizes of the extra dimensions that leads to the
unification of strengths of forces in the framework of the
higher-dimensional theory.

We start our analysis by considering a higher-dimen\-sional formulation
of an $SU(N)$ Yang-Mills theory with matter. We assume that the vector
and scalar particles, i.e. the gauge bosons, $\hat A^\alpha_M$, and
the Higgs field, $\hat \phi$, may propagate
in $4+d$ dimensions, while all the fermionic particles, $\Psi$, are
localized on the 4-dimensional boundary\footnote{Our results apply
also in the case where the fermions live in the bulk.}. 
The $(4+d)$-dimensional action functional, that describes the above
theory and preserves Lorentz invariance, may be written as~\cite{Peskin}
\ba
& & \hspace*{-3mm} S_{4+d}= \int d^4x\,d^dz\,\Biggl\{\nonumber \\
&-& \frac{1}{4}\,\Bigl(\partial_M \hat{A}^\alpha_N-
\partial_N \hat{A}^\alpha_M+\frac{\hat{g}}{\sqrt{\Lambda^d}}\,
C_{\alpha\beta\gamma}\,\hat{A}_M^\beta\,\hat{A}_N^\gamma\Bigr)^2
\nonumber \\[1mm]
&-& \bar{\Psi}_{L,R} \gamma^\mu \Bigl(\partial_\mu
-\frac{i\,\hat{g}}{\sqrt{\Lambda^d}}\,t^\alpha \hat{A}^\alpha_\mu
\Bigr)\,\Psi_{L,R}\,\delta(\vec{z}) \nonumber\\[1mm]
&-& \Bigl|\Bigl(\partial_M 
-\frac{i\,\hat{g}}{\sqrt{\Lambda^d}}\,t^\alpha \hat{A}^\alpha_M
\Bigr)\,\hat{\phi}\Bigr|^2 - \hat{\mu}^2\hat{\phi}^*\hat{\phi}
-\frac{\hat{\lambda}}{2\Lambda^d}\,
(\hat{\phi}^*\hat{\phi})^2  
\nonumber\\[1mm]
&-& 
\Bigl(\frac{\hat{Y}_1}{\sqrt{\Lambda^d}}\,\bar{\Psi}_L\,\hat{\phi}\,\Psi_R
+ \frac{\hat{Y}_2}{\sqrt{\Lambda^d}}\,\bar{\Psi}_L\,\hat{\phi}_c\,\Psi_R
+ h.c. \Bigl)\,\delta(\vec{z}) \Biggr\}\,,
\label{daction}
\ea
where $\hat{g}$, $C_{\alpha\beta\gamma}$ and $t^a$ are the coupling constant,
the structure constants and the generators, respectively, of the $SU(N)$
gauge group while $\hat \mu^2$ and $\hat \lambda$ are the mass and coupling
constant of the Higgs field. In order to render the coupling constants of
the theory dimensionless in $(4+d)$-dimensions, an arbitrary energy scale
$\Lambda$ has been introduced. Note that, in the above,
$M, N=\{t, x_1, x_2, x_3, z_1, z_2,..., z_d\}$,
$\mu, \nu=\{t, x_1, x_2, x_3\}$ and the hat 
denotes $(4+d)$-dimensional quantities.

Next, we assume that the extra $d$ dimensions are compactified over an
internal manifold with the size of every compact dimension being $2L$.
Then, we can Fourier expand the $(4+d)$-dimensional vector and scalar
fields along the compact dimensions in the following way
\be
\hat{\Phi}(x,z) = \hat{\Phi}^{(0)}(x) + \sum_{\vec{n}=1}^\infty\,
\frac{\hat{\Phi}^{(\vec{n})}(x)}{\sqrt{2}}\,\Bigl(
e^{i\frac{\pi\vec{n}}{L}\vec{z}} + e^{-i \frac{\pi\vec{n}}{L}\vec{z}}
\Bigr)\,, 
\label{kk}
\ee
where $\vec{n}=\{n_1, n_2,..., n_d\}$. By performing a
Kaluza-Klein compactification, i.e by using the above field expansion
and integrating over the extra dimensions, the action (\ref{daction})
reduces to an effective 4-dimensional theory. A prominent feature of this
effective theory is its complexity due to extra terms involving
the massive Kaluza-Klein (KK) excitations, $\hat{\Phi}^{(\vec{n})}(x)$,
of all the fields propagating in the extra dimensions, apart from the
usual, massless
zero-modes, $\hat{\Phi}^{(0)}(x)$.  Here, we are only interested in the part of the
effective theory that contains the zero modes of the various fields
and, more specifically, in the relations that hold between the $(4+d)$
and 4-dimensional couplings and masses. These are found to be 
\be
g=\frac{\hat{g}}{(2L \Lambda)^{d/2}}\,, \qquad
Y_{1,2}=\frac{\hat{Y}_{1,2}}{(2L \Lambda)^{d/2}}\,,
\label{gauge}
\ee
\vspace*{-5mm}
\be
\mu^2=\hat{\mu}^2\,, \qquad
\lambda=\frac{\hat{\lambda}}{(2L \Lambda)^d}\,, 
\label{higgs}
\ee
where we have used the following field redefinitions 
\be 
A_\mu^a=(2L)^{d/2}\,\hat{A}^a_\mu\,\,, \qquad
\phi=(2L)^{d/2}\,\hat{\phi}
\label{fields}
\ee
in order to obtain canonical kinetic terms in 4-dimen\-sions.
Note that the fermionic fields being always localized on the
4-dimensional boundary remain unchanged. 

In the framework of the 4-dimensional effective theory, the gauge bosons
and fermions acquire mass through the spontaneous symmetry breaking of
the gauge group. Their masses are given in terms of the vacuum expectation
value of the Higgs field which is also redefined according to
eq.~(\ref{fields}). 
The question, then, arises  whether the mass spectrum
of the 4-dimensional effective theory corresponds to a different one
in the framework of the original (4+d)-dimensional theory. To
answer this question, we consider the masses of the fermions which are
given by the following expression
\be
m_{\Psi_{L,R}}=Y_{1,2}\,\frac{\upsilon}{\sqrt{2}}=
\frac{\hat{Y}_{1,2}}{\Lambda^{d/2}}\,\frac{\hat{\upsilon}}{\sqrt{2}}
=\hat{m}_{\Psi_{L,R}}\,.
\label{masses}
\ee
Similar relations can also be written for the masses of the gauge bosons
and the Higgs field. In all cases, the tree-level masses, that are
generated via the Higgs mechanism in the 4-dimensional theory, remain
unaltered when one embeds this theory in a higher-dimensional one. 

However, this is not the case with the coupling constant $\hat g$
that eventually determines the strength of the gauge
interactions of the theory. The $SU(N)$ gauge group of the
$(4+d)$-dimensional theory (\ref{daction}) could be replaced
by the $U(1) \times SU(2) \times SU(3)$ group with the gauge field
$\hat A^\alpha_M$, the generators $t^\alpha$ and the coupling constant
$\hat g$ standing for each one of the corresponding quantities of the
Standard Model group, i.e. $(\hat{B}_M, \hat{W}^a_M,
\hat{G}^\alpha_M)$, $(Y/2,\tau^a/2, \lambda^\alpha/2)$ and
$(\hat{g}_1, \hat{g}_2, \hat{g}_3)$, respectively. In that case,
each coupling constant and gauge boson of the SM group is redefined
according to eqs.~(\ref{gauge}) and (\ref{fields}), respectively.

By making use of the redefinition (\ref{gauge}) for the gauge coupling
constants, we find that the electric charge $e$ changes as follows
\be
\hat{e} = \hat{g}_1 \cos\hat{\theta}_W = (2L\Lambda)^{d/2}
g_1 \cos\theta_W =(2L\Lambda)^{d/2}\,e\,,
\ee
where we have used the fact that the Weinberg angle $\theta_W$
remains unaltered since its tangent is given by the ratio of the gauge
couplings $g_1$ and $g_2$. The above rescaling of the electric charge
inevitably affects the strength of the electromagnetic interactions in
higher dimensions. The strength of a force is classically defined as the
potential energy of the corresponding interaction between two identical
particles, separated by a distance equal to the particle's Compton
wavelength, compared to the energy of the rest mass of each particle.
As is well known, in 4-dimensions and for two particles with mass
$m$ and charge $e$, the above rule gives the result
\be
\alpha_{EM}=\frac{E_{int}}{E_m}=\frac{1}{mc^2}\,\frac{e^2}
{4\pi\epsilon_0\,(\hbar/mc)}=\frac{e^2}{4\pi\epsilon_0}
\simeq \frac{1}{137}\,,
\ee
where we have set $c=\hbar=1$. 
Generalizing the above definition of the strength of a force in $(4+d)$
dimensions, we obtain
\be
\hat{\alpha}_{EM}=\frac{\hat{E}_{int}}{\hat{E}_m}=
\frac{1}{\hat{m}}\,\frac{(\hat{e}^2/\Lambda^d)}{4\pi\epsilon_0\,
(1/\hat{m})^{1+d}}=\alpha_{EM}\,\Bigl(\frac{m}{M_x}\Bigr)^d\,,
\label{electro}
\ee
where we have used the fact that the masses of the particles do not
change, according to (\ref{masses}), and where we have defined
$M_x \equiv (2L)^{-1}$. The above result clearly 
reveals that the strength of the electromagnetic force changes as the
gauge bosons start ``feeling" the extra dimensions. Moreover, the
strength of the force strongly depends not only on the number and size 
of extra dimensions but also on the mass of the test particle, i.e.
on the energy scale where the measurement takes place. 
It is also worth noting that the auxiliary energy scale $\Lambda$
does not appear in the electromagnetic strength
formula (\ref{electro}).

So far, we have not considered the gravitational interactions. 
We now assume that gravitons may propagate, apart from the usual
4 dimensions, in $d'=\delta + d$ extra dimensions, where $\delta$ is
the number of transverse dimensions, with size $2L'$, felt only by
gravitons. In that case, the $(4+d')$-dimensional action functional  
\be
S_{4+d'}  =  -\frac{M^{2+d'}_{GR}}{16\pi}
\int d^4x\,d^{d'}z\,\sqrt{G_{4+d'}}\,R_{4+d'}
\label{grav}
\ee
reduces to an effective, 4-dimensional Einstein's theory of gravity 
only when the following relation between the sizes of the
extra dimensions and the energy scales of gravity, $M_P$ and 
$M_{GR}$, in 4 and $(4+d')$ dimensions, respectively, 
\be
(2L')^\delta\,(2L)^d\,M^{2+d'}_{GR}=M_P^2 
\label{defmgr}
\ee
holds. The strength of the gravitational interactions also changes
when one introduces extra dimensions for the gravitons. Using the
same rule as above, the strength of the gravitational
interaction between two particles with mass $m$, in 4 dimensions,
is given by the expression
\be
\alpha_{GR}=\frac{E_{int}}{E_m}=\frac{1}{mc^2}\,\frac{G_N\,m^2}
{(\hbar/mc)}=\Bigl(\frac{m}{M_P}\Bigr)^2\,,
\ee
where $G_N\equiv M_P^{-2}$ and natural units, $\hbar=c=1$, have been used
again. The corresponding expression in $4+d'$ dimensions takes the form
\be
\hat{\alpha}_{GR}=\frac{\hat{E}_{int}}{\hat{E}_m}=\frac{1}{\hat{m}}\,
\frac{\hat{G}_N\,\hat{m}^2}{(1/\hat{m})^{1+d'}}=
\Bigl(\frac{m}{M_{GR}}\Bigr)^{2+d'}\,\,,
\label{gravi1}
\ee
where, now, $\hat{G}_N=1/M_{GR}^{2+d'}$. By choosing appropriately the
sizes of the transverse and longitudinal, extra 
dimensions, the higher-dimensional
gravity scale, $M_{GR}$, could be much lower than the 4-dimensional one,
a fact which, subsequently, changes the strength of gravity.

The question we would like to address is the following:
can the gravitational and electromagnetic forces\footnote{Under
the assumption that the electromagnetic, weak and strong
forces ``feel" the same number $d$ of extra dimensions, we may
assume that their strengths remain comparable at every scale even
in $(4+d)$-dimensions.}
have comparable strengths in a world where both gravitons and gauge bosons
feel extra dimensions? If so, 
\be
\hat{\alpha}_{EM}=\hat{\alpha}_{GR} \Rightarrow
\alpha_{EM} \biggl (\frac{m_U}{M_x} \biggl )^d = \biggl (\frac{m_U}{M_{GR}}
\biggr )^{2+d'} \;,
\label{unif}
\ee
where $m_U$ denotes the scale where the unification of strengths takes
place.
Hereafter, for simplicity, we will take $\alpha_{EM} \simeq 10^{-2}$.
Regarding the values $\delta$ and $d$ of the extra dimensions, we may
form four distinct categories :

({\it i}) $\delta=d=0$ . In this case, $M_{GR} \equiv M_P$ and {\it only} at
energies $m_U \simeq 10^{18}$ GeV, i.e. close to the Planck scale, 
the strengths of the forces become equal.

({\it ii}) $\delta \ne 0$, $d=0$ . Here, we assume that only the gravitational
fields can feel extra $\delta$ dimensions. Then, from eq. (\ref{unif}), we
may easily conclude that 
the unification of strengths takes place at a scale
$m_U=(0.2-0.6) M_{GR}$, for $1 \le \delta \le 7$\footnote{Motivated by M-theory,
we assume that $\delta$, $d \leq 7$.}. This means that the strength of
the gravitational force becomes comparable to that of the other forces
only when the measurement takes place near the gravity scale,
independently of where $M_{GR}$ lies. 
The contact with M-theory \cite{witten} is made for
$\delta=1$ and $M_c \equiv (2L')^{-1}=10^{12}$ GeV. Then, the unification
takes place at $m_U \simeq M_{GUT} \simeq M_{GR} \sim 10^{16}$ GeV.
In the case where $M_{GR} \sim 1$ TeV~\cite{ADD}, the hierarchy problem
is {\it removed} by bringing the gravitational scale down to the electroweak
one, $M_W$. However, the energy gap between $M_W$ and the
compactification scale $M_c$, which, for $\delta=2$, 
amounts to 15 orders of magnitude, remains unexplained.

One could resolve the above problem by increasing the number $\delta$ of
extra dimensions accompanied with a small increase in the value of the
higher-dimensional gravity scale $M_{GR}$. For example, allowing
the compactification scale, $M_c$, to be close to the electroweak
one, i.e. $M_c=10^{\pm 1} M_W$, the gravity scale lies in the range
$M_{GR}=(10^5-10^7)$ GeV, for $\delta \ge 6$. In this case, there is no energy
gap between $M_c$ and $M_W$ while the gravity scale is still 
close to the electroweak scale (the ratio $M_{GR}/M_W$, in this case,
is the same as the one of the top quark mass over the down quark
mass, $m_t/m_d \sim 10^4$).

({\it iii}) $\delta=0$ , $d \ne 0$ . In this case, it is the $d$ extra,
longitudinal dimensions that ``open up" for the gauge and scalar fields,
as well as for gravitons, at the scale $M_x$. Substituting
eq. (\ref{defmgr}) into eq. (\ref{unif}) and solving for $m_U$, we find
that the unification of strengths takes place only near the string
scale, i.e. $m_U=M_s \simeq 10^{18}$ GeV, independently of the number $d$ of
extra dimensions. If we take the new gravity scale, $M_{GR}$, to be
the string scale, then, from eq. (\ref{defmgr}), we are led to the
following relation between $M_x$ and $M_{GR}$:
\be
M_x=10^{-2/d} M_{GR} = (0.01 - 0.5) M_{GR} \;,
\label{eq15}
\ee
for $1 \le d \le 7 $. In this case, the compactification
scale $M_x$ is 1 or 2 orders of magnitude smaller than the gra\-vity scale.
A quite interesting result arises for $d=1$. Then, eq. (\ref{eq15})
gives the result $M_{x} \simeq 10^{16}$
GeV, which coincides with the scale of the unification of gauge
couplings, $M_{GUT}$, in the minimal supersymmetric standard model
(according to the power-law mechanism for the unification of
gauge couplings \cite{unification1,ross},
the presence of the extra dimensions does not affect the
unification scale $M_{GUT}$ when $M_x \simeq M_{GUT}$).
The compactification scale, $M_x$, being very close to $M_s$
does not introduce any new energy scale in the theory and is
sufficiently high to suppress the ``dangerous'' baryon number violating
operators. From that point of view, the above arrangement of energy
scales resembles the one that arises in the framework of string
unification~\cite{kaplunovsky} where the compactification scale, $M_x$,
is very close to the string scale, $M_s$. This arrangement is indeed
necessary in order to prevent the gauge coupling $\hat \alpha_{EM}$
from acquiring an unacceptable small value near the unification scale,
a goal which is also accomplished in our case. Although the hierarchy
between the electroweak scale, $M_W$, and the gravity scale, $M_{GR}$,
still remains, the existence of an extra, longitudinal dimension
that ``opens up" just above $M_{GUT}$ may provide a natural explanation
for the energy gap between $M_{GUT}$ and $M_s$.

({\it iv}) $\delta \ne 0$ , $d \ne 0$ . This is the most general 
case where both gravitational and gauge fields feel a number of extra
dimensions. According to eq. (\ref{unif}), the unification of strengths,
now, takes place at the energy scale
\be
m_U^{2+\delta}=\alpha_{EM}\,M_P^2\,M_c^\delta \;.
\label{eq16}
\ee
We make the natural assumption that the ultimate unification of strengths
occurs near the new gravity scale, i.e. $m_U=M_{GR}$. Then, the combination
of eqs. (\ref{defmgr}) and (\ref{eq16}) leads to the same relation
(\ref{eq15}) between the compactification scale $M_x$ and the gravity 
scale $M_{GR}$. On the other hand, the scale $M_c$, associated with the size
of the transverse dimensions felt only by gravitons, can be found
from the following expression:
\be
M_c^\delta= \alpha_{EM}^{-1}\,\frac{M_{GR}^{2+\delta}}{M_P^2}\;.
\label{eq17}
\ee
Note that it is {\it only} the number of transverse dimensions $\delta$ that
determines the compactification scale $M_c$, in terms of $M_P$ and $M_{GR}$,
exactly in the same way that it is {\it only} the number of longitudinal
dimensions $d$ that determines the corresponding expression of $M_x$.
If we, now, choose $M_{GR}=M_s \simeq 10^{18}$ GeV and $d=1$, the interesting
result $M_x \simeq M_{GUT}$ of case ({\it iii}) arises once again. However,
in order for this picture to be viable in the existence of extra, transverse
dimensions for gravitons, the compactification scale $M_c$ should lie very
close to the string scale for every value of $\delta$. Since
$M_c, M_x \geq M_{GUT}$, the unification pattern of gauge couplings is not
affected by the presence of the extra dimensions, either transverse or
longitudinal, and the ultimate unification of all forces takes place
exactly at the string scale. Moreover, the compactification scale $M_c$,
being very close to $M_s$, does not introduce any new energy scale in
the theory. The only other case where this problem is also
resolved is when $M_c = 10^{\pm 1} M_W$, according to the analysis
presented in case ({\it ii}). For $\delta=6$ and $M_c=10$ GeV,
the gravity scale lies at the energy scale $2\times 10^5$ GeV.
Then, for $d=1$, eq. (\ref{eq15}) shows that the unification of strengths
takes place only if $M_x$ is exactly two orders of magnitude
smaller than $M_{GR}$, i.e. $M_x=2$ TeV. Remarkably, this is in
perfect agreement with the proposal of the existence of a single
extra dimension for the gauge bosons with size at the TeV scale
necessary for the explanation of the supersymmetry breaking scale
\cite{antoniadis}. The existence of an extra dimension of this size
would lead to the unification of gauge couplings, through the power-law
mechanism \cite{unification1}, at an energy scale
which is, approximately, one order of magnitude larger than $M_x$
\cite{unification1,ross}. The extra hypothesis, that gravitons may
feel 6 extra, compact dimensions, brings the scale of gravity
down to $200$ TeV, thus, completing the picture of unification
of forces and removing the hierarchy problem (by changing
appropriately the values of $\delta$ and $M_c$, we may easily recover
the case where $M_{GUT}=M_{GR}=M_s=10$ TeV \cite{unification1}).

Finally, let us stress that all the results presented above
can be consistently embedded in a Type I string theory framework. 
Combining and rearranging eqs. (\ref{defmgr}) and (\ref{unif}),
we can write our unification constraint in the form
\be
\Bigl(\frac{m_U}{M_P}\Bigl)^2=\hat \alpha_{EM}\,(m_U\,2L')^{-\delta}\,
(m_U\,2L)^{-d}\,,
\label{string}
\ee
which reduces to the Type I string constraint
$M_s^2 \sim (\alpha_{gauge} M_P \sqrt{V})^{-1}$\cite{unification1}
if we identify $m_U$ with $M_s$, and $\hat \alpha_{EM}$
with $\alpha^2_{gauge}$. The total compactified volume is now
$V=(2L)^d (2L')^\delta$ and a T-duality transformation
$m_U\,r \leftrightarrow (m_U\,r)^{-1}$ needs to be performed,
where $r \equiv 2L, 2L'$.

In conclusion, in this letter, we have demonstrated that the addition
of extra, compact dimensions for gravitons and gauge bosons modify
the strengths of gravitational and gauge interactions and consequently
the scale of their unification. When the gravitons propagate in
$4+\delta$ dimensions, the unification takes place only near the higher
dimensional gravity scale, $M_{GR}$. If we choose $M_c \sim M_W$, no new
scale is introduced in the theory while $M_{GR}$ turns out to be a few
orders of magnitude larger than $M_W$. In the case where the gauge bosons
feel $d$ extra dimensions, the unification occurs only near the string
scale, $M_s$. If $M_{GR} \sim M_s$, $M_x$ turns out to be, for $d=1$, of
the order of the gauge coupling unification scale. Thus, the energy gap
between $M_s$ and $M_{GUT}$ is attributed to the existence of one extra
dimension for the gauge bosons which ``opens up" just above $M_{GUT}$. 
An even more remarkable result arises in the case where both $\delta$
and $d$ are non-zero under the assumption that the gravitons can feel 6
extra dimensions with $M_c=10$ GeV. Then, for $d=1$, $M_x$ has exactly
the right value to explain the size of the supersymmetry breaking scale
with the gravity scale being only 2 orders of magnitude larger than $M_x$.
In the framework of this 11-dimensional theory, the unification of
all forces takes place at $200$ TeV and the hierarchy problem is
completely removed. It is worth noting that all the results derived from
the above analysis, based on a non-renormalizable, effective
field theory for gauge and gravitational interactions, can be
consistently embedded in the framework of string theory unification.

The authors would like to thank K.R. Dienes, H.P. Nilles, R.G. Roberts,
G.G. Ross and K. Tamvakis for fruitful discussions and comments.
A.D is supported from Marie Curie Research Training Grants
ERB-FMBI-CT98-3438. P.K. is grateful to the Particle Physics Group
at Rutherford Appleton Laboratory for the warm hospitality and financial
support.



\vspace*{-5mm}
\begin{thebibliography}{99}


\bibitem{witten} E.~Witten, Nucl. Phys. {\bf B471} (1996) 135; 
P.~Horava and E.~Witten, {\em ibid.} {\bf 460} (1996) 506;
{\it ibid.} {\bf 475} (1999) 94.

\bibitem{Lykken} J.D.~Lykken, Phys. Rev. {\bf D54} (1996) R3693.

\bibitem{ADD} N.~Arkani-Hamed, S.~Dimopoulos and G.~Dvali,
Phys. Lett. {\bf B429} (1998) 263; Phys. Rev. {\bf D59} (1999) 086004; 
I.~Antoniadis, N.~Arkani-Hamed, S.~Dimopoulos and G.~Dvali,
Phys. Lett. {\bf B436} (1998) 257.

\bibitem{RS} L.~Randall and R.~Sundrum, Phys. Rev. Lett. {\bf 83}
(1999) 3370.

\bibitem{antoniadis} I.~Antoniadis, Phys. Lett. {\bf B246} (1990) 377;
I.~Antoniadis and K.~Benakli, {\em ibid.} {\bf 326} (1994) 69;
I.~Antoniadis, K.~Benakli and M.~Quiros, {\em ibid.} {\bf 331} (1994) 313.

\bibitem{unification1}
K.R.~Dienes, E.~Dudas and T.~Gherghetta, Phys. Lett. {\bf B436}
(1998) 55; Nucl. Phys. {\bf B537} (1999) 47.


\bibitem{Peskin} E.A.~Mirabelli and M.E.~Peskin, Phys. Rev. {\bf D58}
(1998) 065002; A.~Pomarol and M.~Quiros, Phys. Lett. {\bf B438} (1998)
255; A.~Delgado, A.~Pomarol and M.~Quiros, Phys. Rev. {\bf D60} (1999)
095008; M.~Masip and A.~Pomarol, {\em ibid} {\bf 60} (1999) 096005;
P.~Nath and M.~Yamaguchi, {\em ibid} {\bf 60} (1999) 116004; 
T.~G.~Rizzo and J.~D.~Wells,  {\em ibid} {\bf 61} (2000) 016007.

\bibitem{kaplunovsky} V.S.~Kaplunovsky, Phys. Rev. Lett. {\bf 55}
(1985) 1036.

\bibitem{ross} D.~Ghilencea and G.G.~Ross, Phys. Lett. {\bf B442}
(1998)~165.



\end{thebibliography}
\end{document}